**Low-temperature Spark Plasma Sintering of fine refractory composite powders core-shell: A case of the powders W@Ni**

A.V. Nokhrin [a, (*)], E.A. Lantcev [a], L.S. Alekseeva [a], N.V. Malekhonova [a], M.S. Boldin [a], Yu.V. Blagoveshchenskiy [b], N.V. Isaeva [b], A.V. Terentyev [b], K.E. Smetanina [a], N.V. Sakharov [a], N.V. Melekhin [a], V.D. Chupriyanova [a]

[a] Lobachevsky University, 603022 Nizhny Novgorod, Russia

[b] A.A. Baykov Institute of Metallurgy and Materials Science of the Russian Academy of Sciences, 119334 Moscow, Russia

e-mail: nokhrin@nifti.unn.ru, malekhonova@nifti.unn.ru

**Abstract**

The mechanisms of fast low-temperature Spark Plasma Sintering (SPS) of W + 10% wt. Ni powders were investigated. The powder compositions were obtained in two methods: (i) by mixing W and Ni powders in a specified ratio (hereinafter referred to as W + Ni); (ii) by Ni deposition on the surface of submicron W particles allowing the formation of particles with a core W – shell Ni structure (hereinafter referred to as W@Ni). To reduce the concentrations of oxygen and oxides, the powders were annealed in hydrogen. The solid-phase sintering was performed at various temperatures (1000-1150°C), pressures (40-80 MPa), heating rates (50-500°C/min), and isothermal holding times (0-20 min). The sintering temperatures corresponded to the onset of intense dissolution of W in Ni. The samples had high relative density and small grain sizes. The activation energy of SPS of the mixed powders was close to the one of the grain boundary diffusion. The key mechanism for the compaction of W@Ni particles in the SPS process is Coble creep. The increasing of the sintering temperature was shown to lead to an increase in the solubility of W in Ni and, consequently, to an increase in the number of secondary $Ni_4W$ particles formed during cooling down. The grain growth has a minor effect on the mechanical properties of the W alloy compared to the changes in its phase composition.

---

[(*)] Corresponding author (nokhrin@nifti.unn.ru)

**Abbreviations:** CMM – Chemical-Metallurgical Method; HEBM – High Energy Ball Milling; LPS – Liquid Phase Sintering; SPS – Spark Plasma Sintering; SEM – Scanning Electron Microscopy; UFG – Ultra-Fine Grained (alloy); XRD – X-ray diffraction (analysis); $d$ – average grain size; $HV$ – hardness; $\rho$ – density; $\sigma_0$ – macroelasticity stress (lattice friction threshold stress); $\sigma_y$ – physical yield strength; W + Ni – powders obtained by mixing, W@Ni – powders with a W core – Ni shell structure obtained by CMM.

## 1. Introduction

Refractory alloys are used extensively high voltage power engineering and nuclear engineering [1-3]. The classical method of manufacturing refractory alloys is liquid-phase sintering (LPS) technology. This process enables the formation of a composite structure in which strong refractory grains are surrounded by a ductile low-melting γ-phase [3]. The good combination of strength, hardness and ductility of refractory tungsten alloys (RTAs) obtained by LPS is primarily determined by the composition and structure of the grain boundaries. A decrease in the fraction of 'ductile' W/γ boundaries and an increase in the fraction of 'brittle' W/W boundaries lead to a worsening in the mechanical properties and performance characteristics of RTAs [4-9].

Currently, the RTAs with a low content of low-melting metal phase (Me = Ni, Fe, etc.) are being investigated extensively as promising materials for structural elements of International Thermonuclear Experimental Reactor (ITER) [10-12]. Consequently, the number of studies dedicated to the development of RTAs for nuclear fusion energy has been increasing rapidly in recent years. Adding a more ductile metal with a small neutron capture cross-section to W contributes to an increased resistance to crack development when RTAs are subjected to plasma pulses. In [11], the additions of 3-5% Ni and Fe were demonstrated to enhance the resistance of W to the impact of powerful deuterium plasma pulses. A significant role of interfacial boundary characteristics in ensuring the high resistance of RTAs to plasma impact was demonstrated in [12]. Modern nuclear power requires high-strength RTAs with high ductility and improved crack resistance. The strain hardening allows improving the ultimate strength of RTAs up to 1600-1650 MPa while simultaneously significant reducing the elongation to failure [3, 13, 14]. Therefore, conventional deformation hardening methods, such as rotary forging and hydroextrusion, are not the most effective techniques for manufacturing thermo- and plasma-resistant elements for ITER.

One of the efficient ways to enhance the mechanical properties of the RTAs is to form an ultrafine-grained (UFG) microstructure [1, 2, 15]. Nano- and ultrafine powders, including powders obtained by high-energy ball milling (HEBM), are often used as starting materials in the production of UFG alloys [16-19]. A large area of grain boundaries in UFG alloys results in a reduction in the thickness of the W/Me interfacial boundaries. This leads to the formation of a large number of the W/W grain boundaries. These boundaries may be areas of brittle microcrack nucleation during deformation [20]. As a result, fine-grained alloys made from mixed powders often have low tensile mechanical properties [21]. A microstructure with a large number of the W/Me interfacial boundaries must be formed to provide increased ductility of UFG RTAs. In this case, the alloy can have both increased strength and ductility. Recent studies of Ni/Ni-W nanocomposites indicate the possibility of achieving high ductility of these refractory materials [22].

A significant amount of research has been dedicated to the topic of LPS of refractory materials [2, 23]. It should be noted that LPS allows the formation of RTAs with a high fraction of the W/Me interfacial boundaries [1, 2]. However, it leads to intensive grain growth and worsening of the mechanical properties.

The use of composite particles with a core W – shell Me structure (hereinafter referred to as W@Me) opens the way for producing materials with a large area of the W/Me interfacial boundaries [24]. Sintering of W@Me powders reduces the number of the W/W boundaries [25]. Coating technologies are actively used to produce W-Cu refractory composites [26, 27]. Technologies for deposition various low-melting metals on refractory metal and ceramic powders have been developed: W@Ni [27-29], W@Ni-Fe [30, 31], WC@Co [32, 33], WC@Ni [33, 34], WC@Ni-Fe-

Cr [35], WC@Cu [36], Mo@Ni [37], $Al_2O_3$@Ni-W [38, 39], SiC@W [40], (Ti,W,Mo,Ta)(C,N)@Co,Ni [41], etc.

One of the effective methods for producing composite refractory powders with a core-shell structure is the chemical-metallurgical method (CMM) of deposition the metal phase from salt solutions [42-44]. An important advantage of the CMM of obtaining the W@Ni particles compared to HEBM method is the ability to avoid significant contamination of the powders with wear products from the mill parts and the grinding media [45]. Powders produced using the HEBM method are often characterized by a significant presence of agglomerates and a non-uniform particle size distribution. An important advantage of the CMM method is the almost complete absence of agglomerates in W@Me powders: the size and shape of the W@Me particles correspond to the size and shape of the initial tungsten powders. The CMM application also prevents alterations in the thermodynamic properties of W powders, formation on internal stresses after HEBM, oxidation of powders, etc. [46-49].

The first studies showed that RTAs sintered from coated submicron powders (~0.3 μm) produced by the CMM method have theoretical density and improved hardness [43, 44]. Application of increased sintering temperatures (1350-1450 °C, 2 h) led to grain growth of up to 2.4-4.6 μm. As a result, the alloys exhibited tensile strength of 950-1050 MPa and hardness of ~3.4 GPa [43, 44]. Improved mechanical properties of RTAs fabricated by the LPS of electrodeposited W@NiFe powders have also been reported [29]. After sintering in argon at 1470 °C, the alloy was characterized by a grain size of 10-15 μm, a flexural strength of 1245 MPa and a hardness of 316-340 $HV_{0.2}$.

It is necessary to avoid intensive grain growth during the sintering of W@Me particles to preserve the maximum volume fraction of the W/Me boundaries in UFG alloy. Grain growth leads to the destruction of thin layers of the fusible metal phase on the surface of the W@Me particles. During intensive grain boundary migration at sintering, this phase concentrates in the triple joints of the grains. An effective method for obtaining fine-grained materials from W@Me particles is Spark Plasma Sintering (SPS). SPS is an advanced version of fast hot pressing [50, 51]. The advantage of SPS is a possibility to reduce the sintering temperature, to reduce the grain growth rate, and to form a fine-grained microstructure due to very short processing times [50, 51].

Refractory alloys produced using the SPS method are primarily damaged along the W/W boundaries during mechanical testing, resulting in low ductility [52]. Local melting of the metal γ-phase in the contact areas of the powders may occur during SPS [51, 53]. This facilitates the uniform distribution of nickel in the RTAs volume and contributes to an increase in the proportion of the W/Ni interfacial boundaries. The effect of pseudo-partial wetting of the W grain boundaries with nickel also leads to the formation of the W/Ni boundaries [54]. In the future, this will make it possible to move to the solution of the urgent problem of developing alloys with extremely low nickel concentration and increased strength and ductility.

A combination of two technologies: (i) the CMM of depositing thin layers of the Ni onto the surface of fine W particles and (ii) SPS is, in our opinion, an efficient approach to obtaining the fine-grained RTAs with a large area of interfacial boundaries. By preliminary forming a system of viscous W/Ni boundaries on the particle surfaces and preserving it during fast solid-phase sintering, one can obtain a material with enhanced strength and ductility simultaneously. The efficiency of the combined use of these two methods has been demonstrated previously in the development of UFG ultra-low-Co solid alloys [55] and composite ceramics YAG:Nd + Me (Me = Mo, Ni, W) [56, 57].

The aim of the present study was to investigate the microstructure of RTAs alloys fabricated using the solid-phase SPS method. The main focus of the present study was on the challenge of low-temperature rapid sintering of W@Ni powders. The alloys obtained by mixing W and Ni powders followed by SPS were used as the reference. Additional factors affecting (i) density, (ii) solubility of W in the Ni-based γ-phase, (iii) content of viscous γ-phase, primarily depending on the intensity of $Ni_4W$ intermetallic precipitation, (iv) amount of the W oxides, and (v) grain growth were analyzed further: annealing temperatures of the powders in hydrogen, heating rates, sintering temperatures and times as well as the magnitude of the pressure applied.

## 2. Materials and methods

This study focused on the W + 10 wt.% Ni alloys with the theoretical density $\rho_{th}$ = 17.065 g/cm³.

The submicron commercial α-W powders with the Fisher particle sizes of 0.8 μm were utilized to prepare the powder compositions (manufacturer: Kirovgrad Hard Alloys Plant, JSC, Russia). The α-W powders exhibited no significant number of large agglomerates (Fig. 1a). The fine α-W particles had a faceted morphology (Fig. 1b). The oxygen concentration in the α-W powders was 0.65% wt. Such an oxygen concentration corresponds to ~4% wt. $WO_2$. The XRD studies did not detect the presence of the W oxides in the powders. This result indicates the W oxide to be amorphous and to be located on the surfaces of the W particles. Such thin nanolayers of amorphous tungsten oxide cannot be detected by X-ray phase analysis.

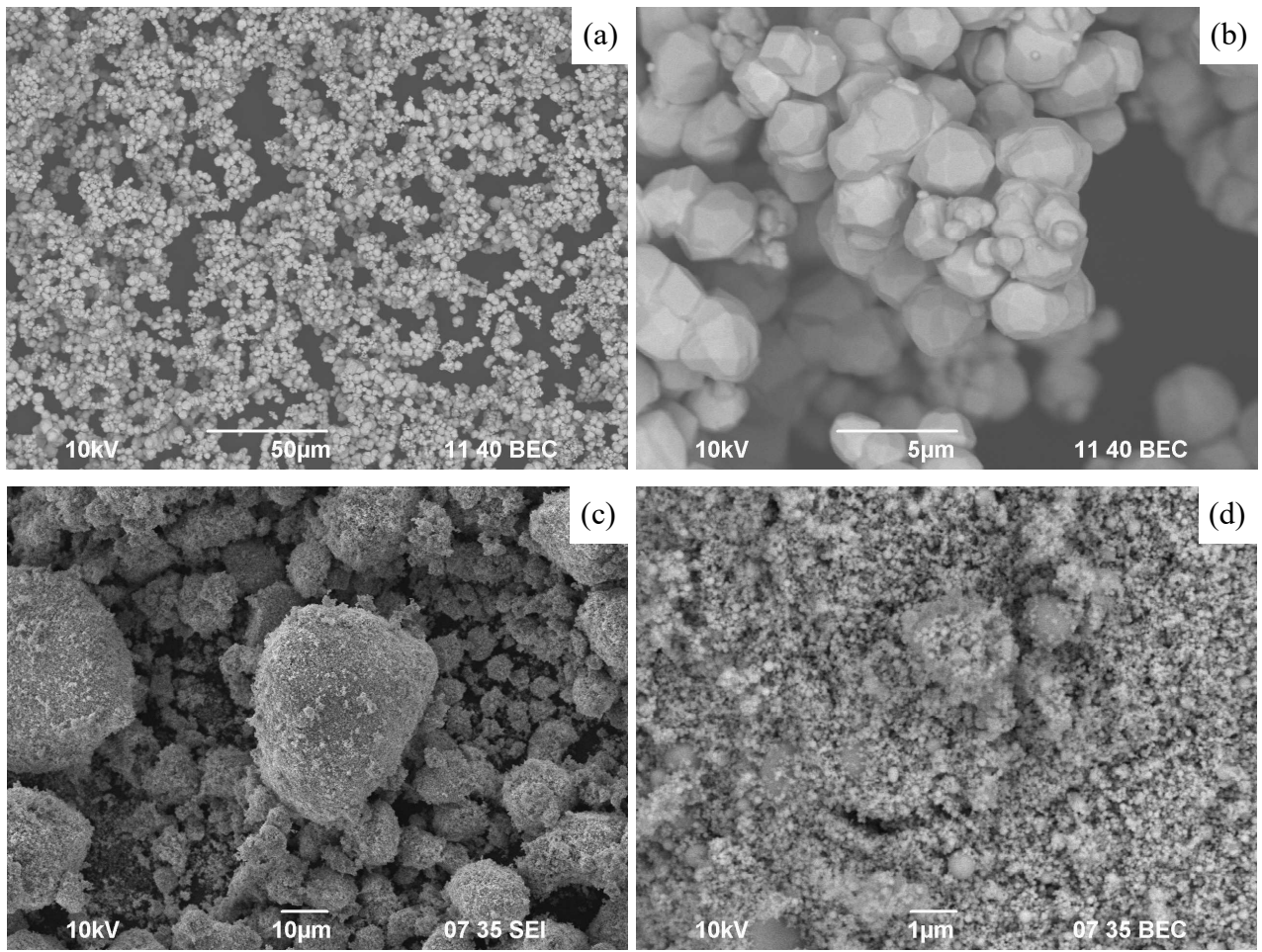

**Fig. 1.** SEM images of W (a, b) and Ni (c, d) powders

The powder compositions of W + 10% Ni were prepared using two methods: the method of mixing W + Ni powders (Method #1) and the CMM of obtaining W@Ni powders (Method #2).

Method #1. The submicron W powders (Fig. 1a, b) and the Ni nanopowders (manufacturer: Advanced Powder Technologies, JSC, Tomsk, Russia) (Fig. 1c, d) were mixed in a ball planetary mill Fritsch Pulverisette in isopropyl alcohol medium. The tungsten carbide-based grinding bodies and hardware were used for mixing. Prior to mixing, the powders were subjected to ultrasonic treatment in alcohol for 15 min. Mixing was performed for 5 h at the rotation speed of 150 rpm. The mixing process aimed to break the Ni agglomerates present in the powders (Fig. 1c) and ensure uniform distribution of Ni particles in the W powder. The mixing parameters were selected on the base of the results of previous investigations (see [17, 19]).

To assess the homogeneity of the mixture, the obtained W + Ni powder mixtures were pressed at room temperature in a steel mold with a diameter of 20 mm. Occasionally, the resulting blank was slightly heated to enhance its strength. Samples for analysis were taken from different areas of the pressed blank. The results of X-ray phase analysis and EDS microanalysis were analyzed for five samples. In some cases, the number of samples was increased to seven. For X-ray studies, the samples were ground using an agate mortar. The powder blank was considered homogeneous if the results of X-ray phase analysis and EDS analysis of samples taken from different areas of the pressed blank were consistent.

Method #2. The α-W powder was added to a beaker containing a solution of Ni hexahydrate $NiCl_2 \cdot 6H_2O$ in alcohol in proportions corresponding to the desired composition. The Ni salts were first dissolved in alcohol at 70°C for at least 20 min. Mixing was performed using RW 20 IKA stirrer. The resulting mixture was dried at 150°C with constant stirring until a dry residue was obtained. Subsequently, it was annealed in a tubular furnace Nabertherm® RS 120/750/13 in the following regime: (i) heating up to 350°C in $H_2$ flow, holding for 30 min, (ii) further heating up to $T_{H2}$ = 400, 500, 600, or 750°C, and holding at the temperature specified for 3 h, (iii) followed by cooling down the silica reactor under an Ar flow. Before hydrogen annealing, the powders were ground in a mortar to eliminate the large agglomerates.

It is important to note that the tungsten powders were not preliminary annealed in hydrogen to reduce oxygen concentration and remove nanolayers of amorphous tungsten oxides $WO_x$. Therefore, the powders synthesized by Method #2 had a structure core W – intermediate amorphous layer $WO_x$ – shell Ni.

The powders obtained by Methods #1 and #2 will be referred to as Powders #1 and #2, respectively. The W alloys manufactured from Powders #1 and #2 will henceforth be denoted as Alloys #1 and #2, respectively.

For further removal of the agglomerates, the powders were subjected to wet milling in a Retch PM-400 planetary mill (grinding bodies and hardware made of tungsten carbide-based alloy), followed by vacuum annealing in Thermotechnik®-ML ESKVE-320 GM13 furnace at 700°C.

The preliminary cold pressing of cylindrical samples weighing 8 g of 12.8 mm in diameter and of 5 mm in height, it was conducted using a steel mold. The optimal uniaxial pressure was determined by investigating the density and homogeneity of compacts made from Powder #1. The compacts were produced using the pressures of 30, 50, and 70 MPa, then sintered in hydrogen at the temperatures $T_{H2}$ = 700, 900, 1000, 1100, 1200, and 1300°C. As shown in Fig. 2, the densities of the compacts made with the pressures of 50 and 70 MPa were nearly identical after heating up to 1000 °C. Since some cracking was observed in the compacts made with the pressure of 70 MPa, the pressure of 50 MPa was selected as the optimal for the cold pressing.

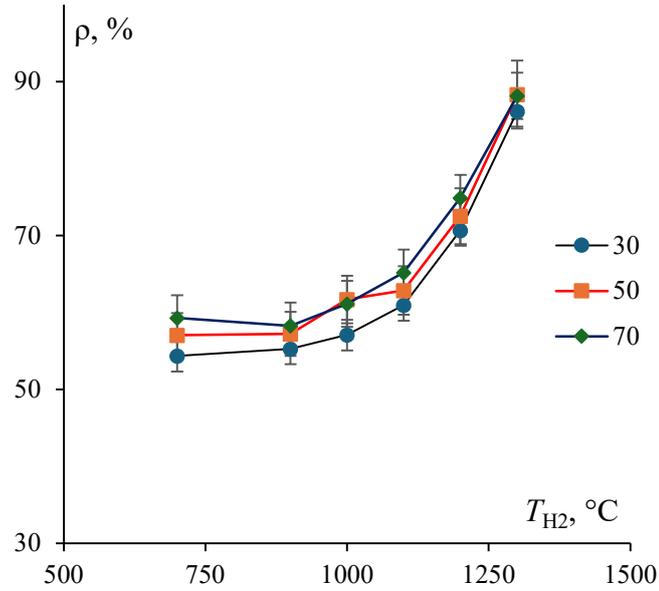

**Fig. 2.** Dependence of the density of W + 10% Ni compacts on the temperature of hydrogen annealing

Sintering was performed in the graphite molds with the internal diameter of 12.8 mm. To compensate for the thermal expansion of the samples and to increase their density and adherence to the inner surface of the graphite mold, the samples were wrapped in thin graphite foil. Graphite foil helps to reduce oxygen concentration during the heating of nickel powders [58]. SPS of the compacts was performed using Dr. Sinter® model SPS-625 setup. The samples were sintered in the continuous heating mode with a specified heating rate ($V_h$) up to the sintering temperature ($T_s$) followed by cooling down in free mode. Heating was achieved by passing the 12 ms electric current pulses through the graphite mold with the 2-ms pauses. The durations of holding at the sintering temperature $T_s$ were 0, 5, or 20 min. The uniaxial pressure ($P$) was applied to the samples simultaneously with the start of heating. The sample temperature was determined using CHINO® IR-AH optical pyrometer focused onto the surface of the graphite mold. The uncertainty of the temperature measurement was ± 20 °C. The effective powder shrinkage ($L_{eff}$) was measured using Futaba® Pulscale™ SMM151A dilatometer built in Dr. Sinter® model SPS-625 setup. To account for the contribution of thermal expansion of the die mold and to determine the true powder shrinkage ($L$), an experiment was conducted by heating the empty molds. The powder shrinkage rate ($S$) was calculated using a linear approximation: $S = \Delta L/\Delta t$.

The surface of the sintered samples was subjected to waterjet cleaning to remove graphite residues. Afterward, the surface of the samples was mechanically ground to remove the carburized layer that results from the interaction of the material with the graphite mold [51, 59]. This is necessary to avoid embrittlement of alloys caused by grain boundary segregation of carbon and oxygen [2, 3, 60].

The effects of the sintering temperature, the heating rate, and the pressure applied on the density and characteristics of the W + 10% Ni alloys sintered from Powder #1 were investigated. This allowed determining the best characteristics for Alloys #1, which were further compared with the ones of Alloys #2 manufactured from the W@Ni powders.

The density of the samples ($\rho$) was measured by hydrostatic weighing using Sartorius® CPA laboratory balance. The uncertainty of the density measurement was ± 0.01 g/cm³.

The powders and the sintered samples were investigated by scanning electron microscopy (SEM) using JEOL® JSM-6490 instrument equipped with Oxford Instruments® INCA 350 EDS microanalyzer. To study the structure, the sintered samples were mechanically polished using a suspension of aluminum oxide nanoparticles with a particle size of 50 nm. The average grain size ($d$) was determined by the chord method using GoodGrains software. Based on the analysis of SEM images (BEC contrast), the contiguity coefficient ($C_{SS}$) was calculated using the formula: $C_{SS} = 2N_{WW}/(2N_{WW} + N_{WNi})$, where $N_{WW}$ is the number of grain W/W boundaries, and $N_{WNi}$ is the number of interphase W/Ni boundaries [6, 9]. The values of $N_{WW}$ and $N_{WNi}$ were determined by superimposing an X-Y scale grid and counting the number of line intersections. In fine-grained alloys #2, the thickness of the interphase W/Ni boundaries was very small, which led to errors in calculating the $C_{SS}$ coefficient. Additional verification of the number of W/W and W/Ni boundaries was carried out on sample fractures using the SEM method (BEC contrast) at high magnification.

X-ray diffraction (XRD) phase analysis was carried out using Haoyuan Instrument DX-2700BH diffractometer (CuK$_\alpha$ radiation, scanning step 0.04°, and exposure time 2 s). Qualitative phase analysis was conducted with the DIFFRAC.EVA software (Bruker, Germany) using data from the PDF-2 database (ICDD, 2012) and cif-files (ICSD, 2015). For clarity in presenting the results, all powder and sintered sample X-ray diffraction patterns are plotted with a constant vertical offset of 250 pulses relative to each other. The intensity axis (Y-axis) scale is consistent across all X-ray diffraction patterns, accounting for this offset.

The Vickers hardness (*HV*) of the sintered samples was measured using Qness® A60+ hardness tester with a load of 10 kg. The average *HV* measurement uncertainty was ± 0.1 GPa. At least ten hardness measurements were taken in the central part of the sample, and at least ten hardness measurements were taken at the edges of the sample (five measurements on each of two opposite edges). All measurements were conducted under identical conditions (using the same load).

### 3. Results

3.1 Powders W + Ni obtained by mixing (method #1)

Fig. 3 shows SEM images of Powder #1 at various magnifications. Fig. 3a shows that the powders contained some individual agglomerates, the size of which varied from 5 to 20 μm. The average sizes of the powder particles in the initial state (after mixing) were close to 0.8 μm (Fig. 1b) that corresponded to the vendor's specifications (Kirovograd Hard Alloys Plant, JSC).

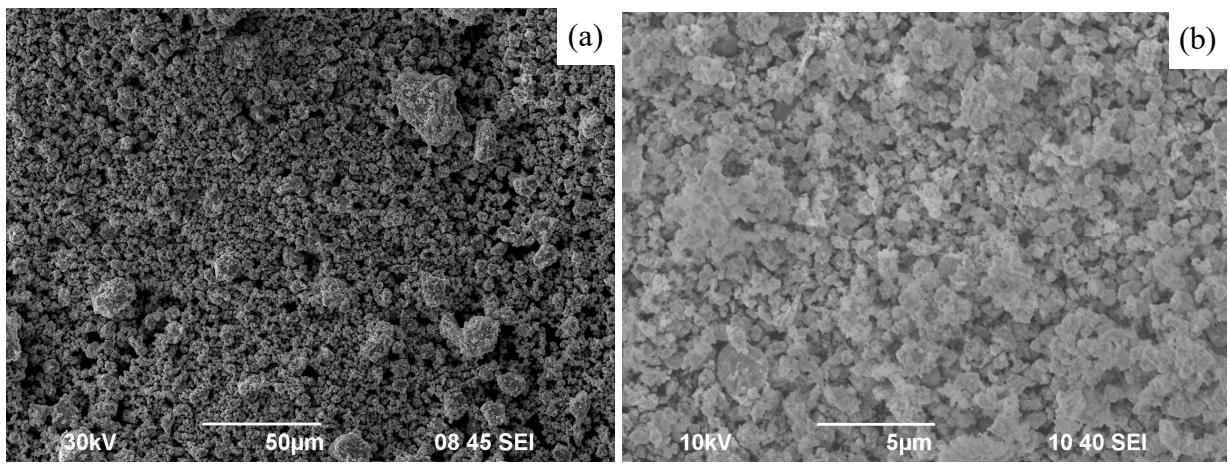

**Fig. 3.** Powders W + 10% Ni obtained by mixing (Method #1). SEM

The XRD results indicated the powders to be the two-phase ones. The XRD curves manifested the peaks from W (PDF #00-004-0806, ICSD #76151) and Ni (PDF #00-004-0850, ICSD #646089) only (Fig. 4). In the XRD curves from the powders annealed at 700°C, most remaining XRD peaks can be described by the structural type of Ni with an enlarged elementary cell (≈ 3.568 Å) (Fig. 4b). Such a lattice parameter in W-Ni alloys corresponds to the ratio Ni:W = 9:1 (% at.) according to [61]. Further, such a compound (γ-phase) will be denoted in the XRD curves as $Ni_{0.93}W_{0.07}$. Additionally, an additional peak was observed at 2θ ≈ 29°, which does belong neither to W nor to Ni. In the XRD curve of the powder annealed at 900°C, the peaks related to W or to $Ni_4W$ phase (PDF #03-065-2673, ICSD #105452) were observed (Fig. 4a). In the XRD curves from the powders annealed at higher temperatures, there was no separation of the (002) and (301) XRD peaks at 2θ ≈ 51°.

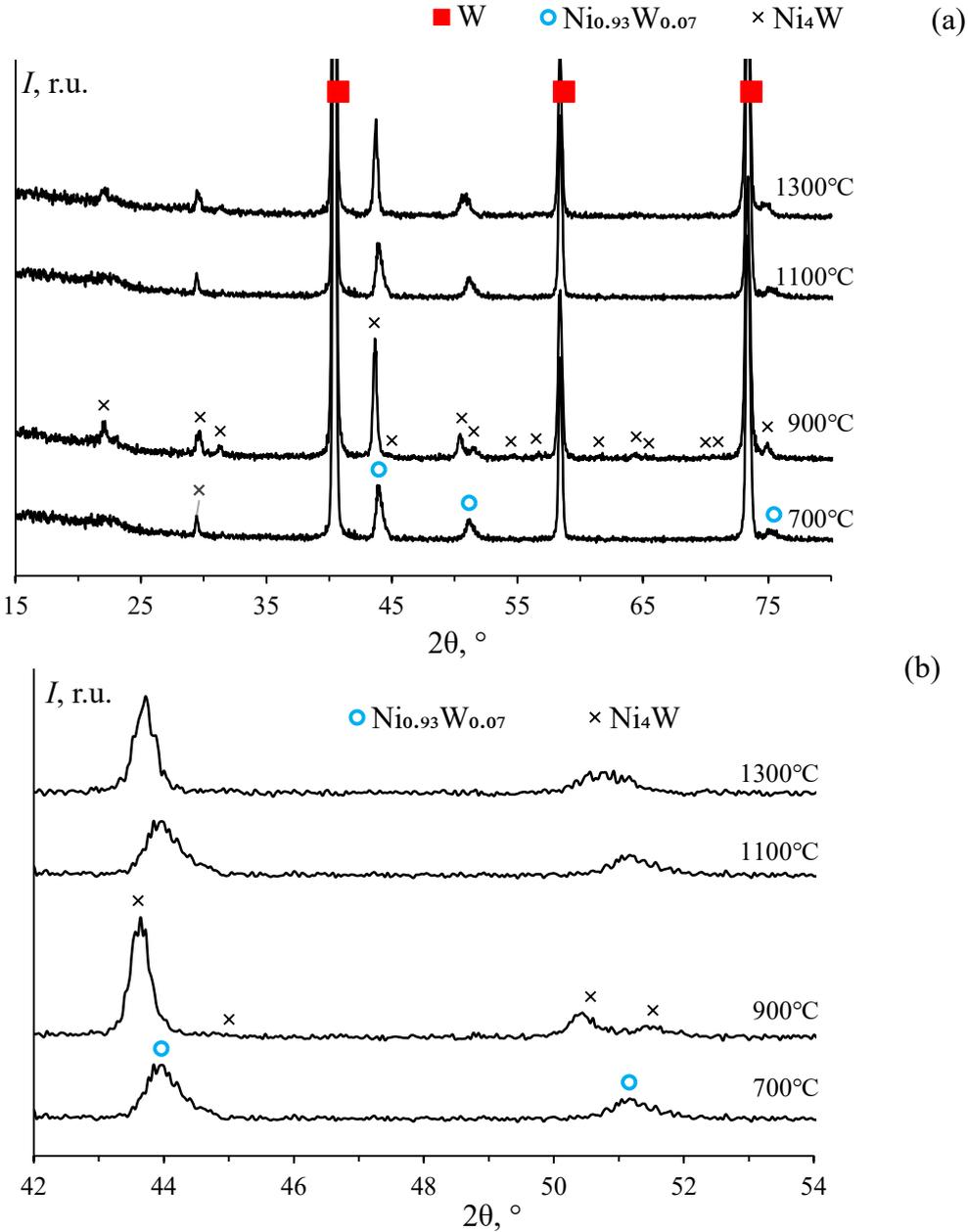

**Fig. 4.** XRD curves of Powder #1: (a) overall view; (b) fragment of the XRD curves 2θ = 42-54°. The temperatures of annealing in hydrogen are indicated on the XRD curves

Note that the Ni$_4$W phase forms usually as a result of tungsten oxide decomposition in the presence of Ni and hydrogen [62]. Therefore, one can assumed the process of Ni$_4$W phase formation to indicate the presence of minor additions of WO$_2$ in the initial W powders. An increased oxygen concentration in the investigated W powders (~0.65% wt.) supports this assumption indirectly.

The analysis of the XRD results indicated the annealing temperature to affect the W concentration in Ni. At 700°C, the W concentrations in Ni were sufficient for the appearance of a superstructure with a period of ≈4 Å leading to the disruption of the cubic structure of Ni and the appearance of a XRD peak at $2\theta \approx 29°$. Further increasing of the W content from ~10% at. up to ~20 % at. in the powder annealed at 900°C resulted in the formation of the Ni$_4$W intermetallic compound with tetragonal symmetry. Further increasing of the W concentration led likely to distortion of the Ni$_4$W tetragonal cell resulting in a shift of the (002) and (301) XRD peaks at $2\theta \approx 51°$.

Fig. 5 presents the shrinkage temperature curves $L(T)$ for Powder #1. For the ease of comparison of the samples with different initial masses, the $L(T)$ curves were transformed into densification temperature curves $\rho/\rho_{th}(T)$ according to the procedure described in [63]. Fig. 5 illustrates that regardless of the preliminary treatment regime of the powders or the SPS mode, the curves $\rho/\rho_{th}(T)$ had conventional three-stage character: the low-temperature stage with a low sintering rate (Stage I, $T \geq T_1$), the intensive powder densification stage (Stage II, $T_1 < T \leq T_2$), and the high-temperature stage, which the powder shrinkage rate is low within (Stage III, $T \geq T_2$).

Annealing in hydrogen at $T_{H2}$ ~ 700°C let to an increase in the shrinkage rate at Stage II and to an increase in the temperatures $T_1$ but had no effect on the maximum shrinkage $\rho/\rho_{th}$ at ~1100°C (Fig. 5a). Increasing the annealing temperature $T_{H2}$ led to a decrease in $\rho_{max}/\rho_{th}$ that is likely due to a reduction in the amount of ductile γ-phase Ni$_{0.93}$W$_{0.07}$ and to an increase in the amount of Ni$_4$W (Fig. 4a). Increasing the heating rate from 50 up to 500 °C/min led to a decrease in the shrinkage rate and to a slight increase in the temperatures $T_1$ and $T_2$ (Fig. 5b). The general shape of the $L(T)$ curves did not change with increasing heating rate or pressure applied (Fig. 5b, c). Increasing the pressure from 40 up to 80 MPa led to a monotonous increase in the shrinkage at Stages I and II but did not have a significant effect on the temperatures $T_1$ and $T_2$ (Fig. 5c).

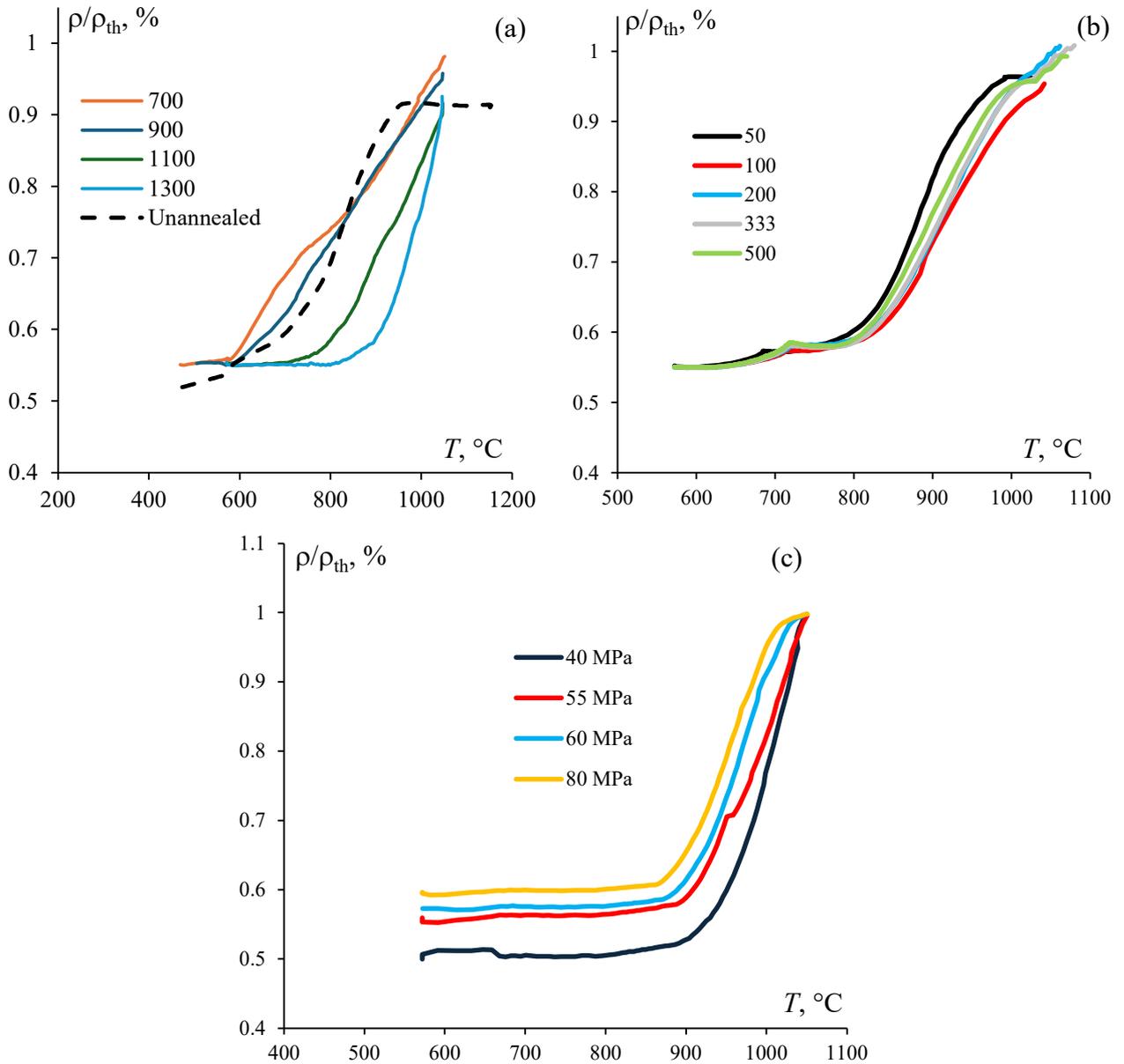

**Fig. 5.** Temperature curves $L(T)$ of Alloy #1: (a) effect of hydrogen preliminary annealing at $T_{H2}$ = 700, 900, 1100, and 1300°C; (b) effect of heating rate $V_h$ (50, 100, 200, 333, 500°C/min); (c) effect of the pressure applied $P$ (40, 55, 60, 80 MPa)

The results of the investigation of the microstructure and properties of RTAs are presented in Table 1. Table 1 shows that increasing the sintering temperature $T_s$ from 1075 up to 1150°C led to an increase in the relative density of the alloys sintered from the unannealed powders from 92.1 to 95.4%. Increasing the heating rate $V_h$ from 50 up to 500°C/min and increasing the $P$ from 40 up to 80

MPa did not lead to any significant change in the alloy density. Hydrogen annealing of the powders at 700°C led to an increase in the density of alloys whereas the $WO_2$ particles were absent in the alloys. Increasing the hydrogen annealing temperature of powders led to the formation of $Ni_4W$ intermetallics and to a decrease in the density of the RTAs sintered at 1050 and 1100°C.

Table 1. Characteristics of the W + 10% Ni alloys manufactured from Powder #1

| $T_{H2}$, °C | SPS mode | | | | Alloy characteristics | | | |
|---|---|---|---|---|---|---|---|---|
| | $V_h$, °C/min | $T_s$, °C | $t_s$, min | $P$, MPa | $\rho/\rho_{th}$, % | XRD [1] | $d$, μm | $HV$, GPa |
| – | 50 | 1075 | 5 | 80 | 92.1 | $WO_2$ | 0.5 | 8.7 [2] |
| | | 1100 | | | 91.9 | | 0.5 | 7.3 |
| | | 1125 | | | 91.6 | | 0.7 | 6.9 |
| | | 1150 | | | 95.4 | | 1.0 | 6.0 |
| 700 | 50 | 1000 | 5 | 80 | 97.9 | $Ni_4W$ | 0.8 | 5.5 |
| | | 1025 | | | 99.5 | | 1.0 | 5.4 |
| | | 1050 | | | 99.7 | | 1.0 | 5.4 |
| | | 1075 | | | 99.7 | | 1.1 | 5.3 |
| | | 1100 | | | 99.8 | | 1.1-1.2 | 5.3 [2] |
| 700 | 50 | 1050 | 0 | 80 | 97.1 | $Ni_4W$ | 0.8 | - |
| 900 | | | | | 95.8 | | 0.8 | - |
| 1100 | | | | | 91.6 | | 0.9 | - |
| 1300 | | | | | 90.8 | | 1.0 | - |
| 700 | 50 | 1100 | 0 | 80 | 99.4 | $Ni_4W$ | 1.0 | - |
| 900 | | | | | 99.2 | | 1.0 | - |
| 1100 | | | | | 99.1 | | 1.1 | - |
| 1300 | | | | | 97.8 | | 1.2 | - |
| – | 50 | 1030 | 0 | 80 | 96.8 | $WO_2$ | 1.0 | 10.0 [2] |
| | 100 | | | | 95.4 | | 0.8 | 7.3 |
| | 200 | | | | 95.8 | | 0.7 | 7.1 |
| | 333 | | | | 96.3 | | 0.5 | 6.9 |
| | 500 | | | | 96.5 | | 0.5 | 6.6 |
| 1050 | 50 | 1030 | 0 | 40 | 99.5 | $Ni_4W$ | 0.8 | 5.5 |
| | | | | 50 | 99.5 | | 0.8 | 5.5 |
| | | | | 55 | 99.6 | | 0.8 | 5.4 |
| | | | | 60 | 99.8 | | 0.8 | 5.4 |
| | | | | 80 | 99.8 | | 0.8 | 5.5 |

[1] The presence of the impurity phases in the W alloy (besides W and Ni) is indicated.

[2] The cracks formed during the hardness measurements.

The results of investigations of the mechanical properties have shown the samples of fine-grained W + 10% Ni alloys to demonstrate high values of hardness $HV$ (Table 1). Severe plastic deformation of the surface was observed near the Vickers pyramid imprints during the hardness measurement (Fig. 6a). The mechanical properties of the fine-grained Alloys #1 were similar to those of the UFG alloys obtained by SPS from the nanopowders [17, 19]. The best mechanical properties ($HV$ = 10 GPa) were observed for the alloy sintered at 1030°C with the heating rate $V_h$ = 50°C/min at the pressure $P$ = 80 MPa. It is important to note that the samples made from the unannealed powders demonstrated increased brittleness; the cracks formed on the surfaces of such samples during the hardness measurements (Fig. 6b). In the unannealed samples with increased hardness, the cracks formed without any plastic deformation of the surfaces (Fig. 6c). The cracks propagated linearly that indicated a transgranular fracture. The samples made from the $H_2$-annealed powders manifested an increased density but lower values of $HV$. In our opinion, this is because annealing in $H_2$ reduces the amount of $WO_2$ particles, which inhibit the grain growth and plastic deformation of the RTA. As a result, the alloys made from the annealed powders have larger grain sizes than the ones made from unannealed powders (Fig. 7) and reduced hardness (Table 1).

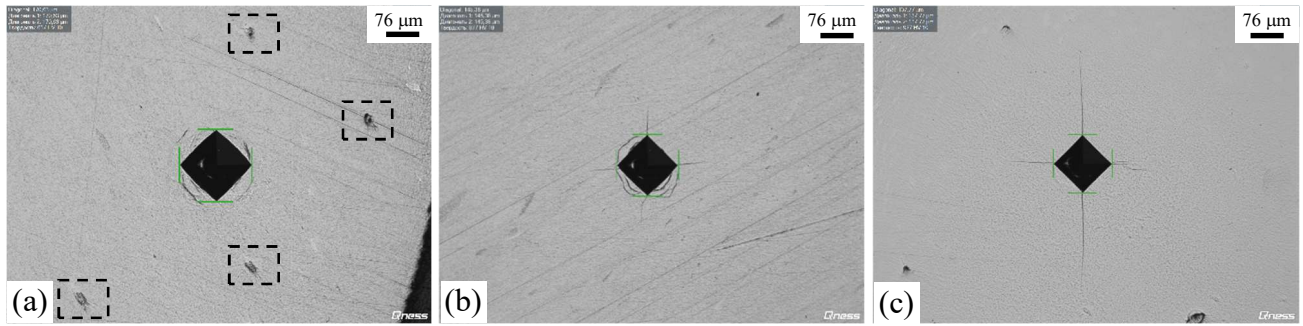

**Fig. 6.** Photographs of Vickers pyramid imprints on the surfaces of the RTA samples: (a) plastic deformation without crack formation; (b) plastic deformation with crack formation; (c) crack formation without plastic deformation

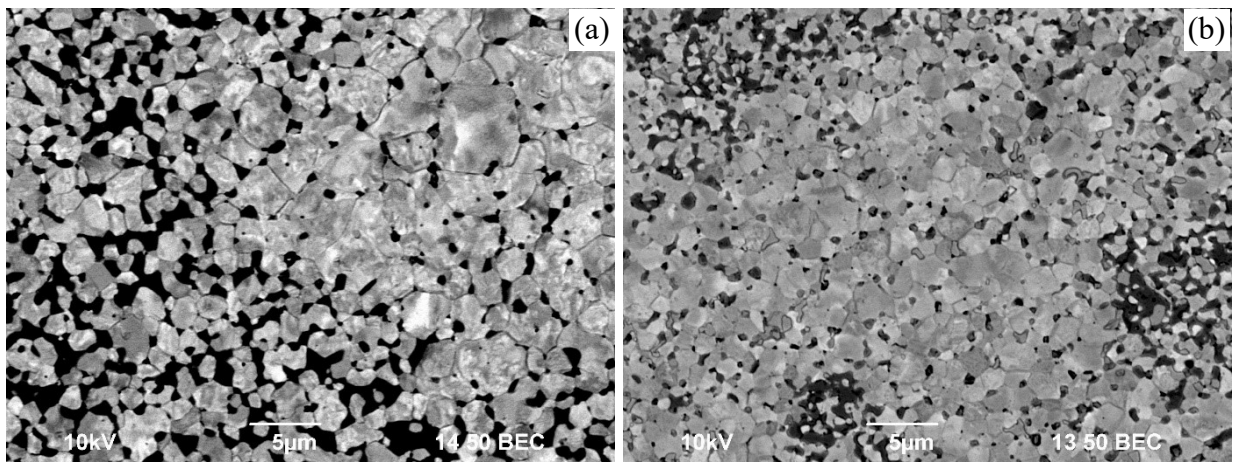

**Fig. 7.** Typical microstructure of Alloy # 1. Sintering temperature 1100°C. SEM

The Alloys #1 exhibited fairly nonuniform fine-grained microstructure. The areas with both decreased and increased Ni contents were observed on the ground surface (Fig. 7). The nonuniform distribution of Ni in the sintered W alloys was presumably due to its uneven distribution in the initial powder compositions as well as the agglomeration of the W powders. At the macroscopic level, individual large inclusions (Fig. 6a) are visible on the ground surface, which are the agglomerates of the Ni particles. In Fig. 6a, the Ni agglomerates on the ground surface are marked by the dashed lines.

3.2 W@Ni powders (Method #2)

Fig. 8 presents the SEM images of the W@Ni particles obtained by CMM. The W@Ni powders had submicron particle sizes and a rounded shape. The large agglomerates were almost absent in the powder mixtures. There were no traces of external impurities (tungsten carbide) detected in the powders. At high magnifications, individual faceted W particles were observed in the SEM images (marked by yellow dashed lines in Fig. 8b), the volume fraction of such particles did not exceed 5%[1]. The presence of such faceted particles indicates that it was not possible to distribute Ni uniformly throughout the powder during the synthesis. Therefore, the powder #2 investigated was a mixture of W@Ni + α-W.

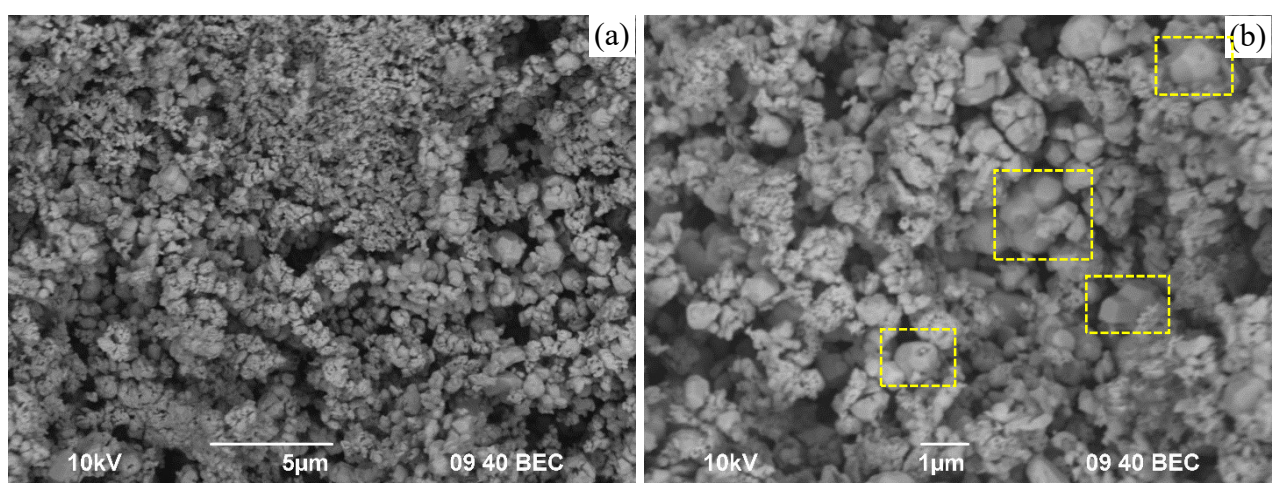

**Fig. 8.** SEM images of W@Ni powders obtained by CMM and hydrogen annealed at 600 °C

The analysis of the XRD curves presented in Fig. 9 shows that the W@Ni powders contain α-W (PDF #00-004-0806, ICSD #76151), Ni (PDF #00-004-0850, ICSD #646089), and the $WO_2$ impurity phase (PDF #04-003-5856). The intensities of the XRD peaks attributed to the $WO_2$ phase decreased with increasing annealing temperature in $H_2$. After hydrogen annealing at 600 °C, only one XRD peak $WO_2$ at $2\theta \sim 26.5°$ was observed, the intensity of which exceeds the background level slightly (Fig. 9). It is interesting to note the splitting of the Ni peaks, which is characteristic of two stoichiometric phases usually. Since the position of the second peak did not change with increasing hydrogen annealing temperature (as would be expected in the case of the formation of a solid solution of W in Ni). The appearance of the second peak is most likely associated with the formation of the $Ni_{0.93}W_{0.07}$ intermetallic (PDF #04-011-9054). At higher hydrogen annealing temperature (750°C), the powders were two-phase ones and consisted of W and the $Ni_4W$ intermetallic (PDF #03-065-2673, ICSD #105452). Using XRD, no Ni particles necessary for forming 'ductility' W/Ni interphase boundaries were detected. Therefore, the powders annealed at 400-600°C were used for further studies.

---

[1] The volume fraction of faceted W particles was determined by calculating the area occupied by these particles relative to the total surface area under study. This calculation was performed using the GoodGrains software. At least five different areas were analyzed, each containing a minimum of 150 tungsten particles.

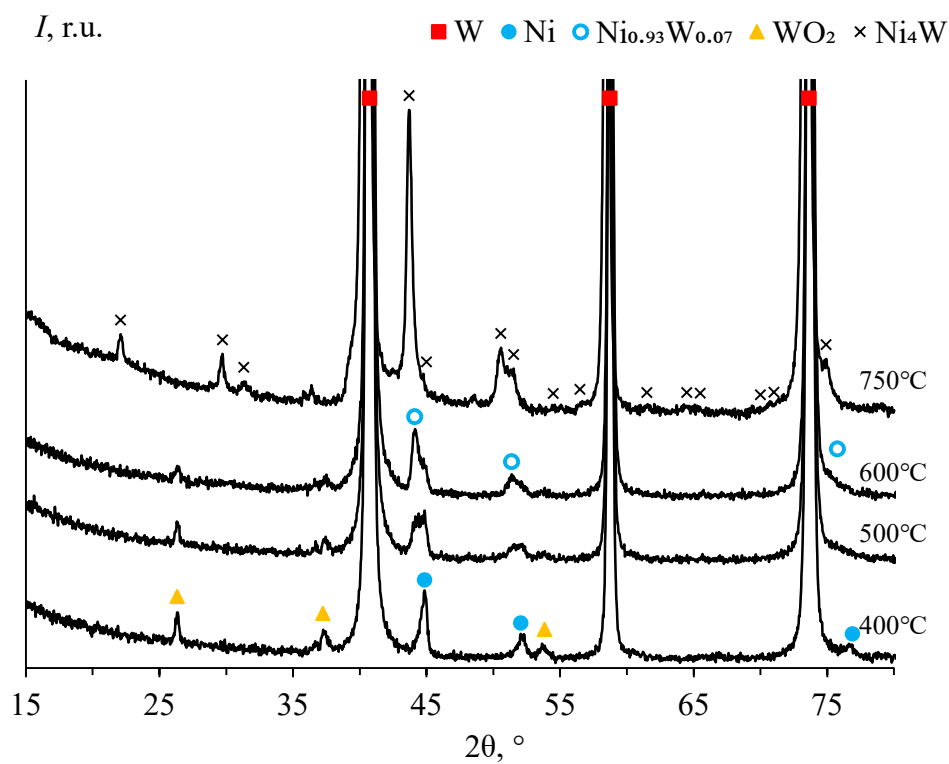

**Fig. 9.** XRD curves of the W@Ni powders after hydrogen annealing at 400, 500, 600, and 750°C (3 h)

Fig. 10 presents the temperature curves of shrinkage $L(T)$ and shrinkage rate $S(T)$ for the W@Ni powders subjected to reductive hydrogen annealing at various temperatures. The $L(T)$ curves exhibited the conventional three-stage character [64]. It can be seen from the $L(T)$ curves presented in Fig. 10a that increasing the hydrogen annealing temperature led to a slight decrease in the shrinkage and shrinkage rate of the powders. The intensive shrinkage of the W@Ni powders takes place at 980-1000°C. The maximum shrinkage rate $S_{max} = (1.7 - 1.8) \times 10^{-2}$ mm/s of W@Ni powders was achieved at 840-850°C.

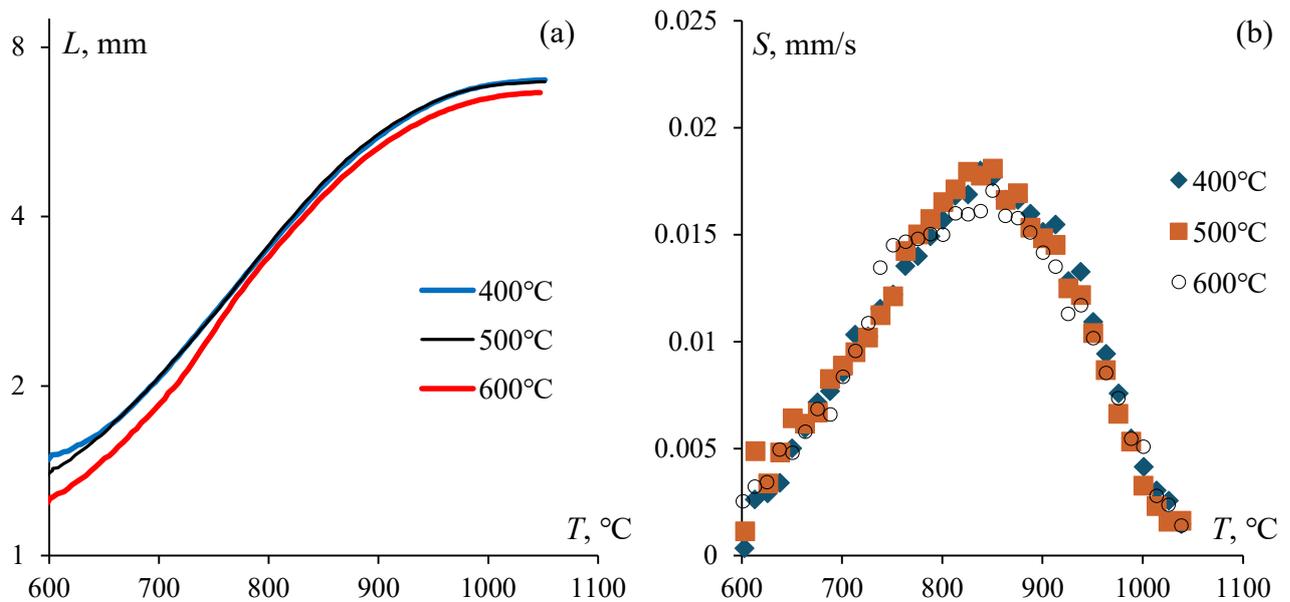

**Fig. 10.** Temperature curves of shrinkage $L(T)$ (a) and shrinkage rate $S(T)$ (b) of W@Ni powders

The densities of the alloys sintered by heating up to 1050°C without isothermal holding ($t_S = 0$ min) were: 99.5, 99.7, and 100%, respectively, for the powders annealed at 400, 500, and 600°C. Increasing the holding time at the sintering temperature (1050°C) led to an increase in the densities of alloys sintered from the annealed powders #2 up to 99.9-100%. Thus, the densities of the Alloys #2 were close to the theoretical one.

The surfaces of the samples interact with carbon intensively during sintering in the graphite molds [59]. To analyze the effect of carburization on the characteristics of the alloys, the sintered samples were cut longitudinally, then each surface was polished mechanically and the phase composition, microstructure, and microhardness of the surface and central layers of the W + 10% Ni samples were investigated.

The results of the XRD investigations revealed the W, Ni, and $WO_2$ phases as well as the $\eta_1$-phase $Ni_3W_3C$ (PDF #01-078-5006, ICSD #166814) in the surface layers of all samples regardless of the annealing mode of the powders (Fig. 11). The formation of impurity phases in the Ni-W-C system was previously discussed during the sintering of mechanically activated Ni-W nanopowders [65]. Lowering the sintering temperature made it possible to avoid the intensive formation of $M_{12}C$ и $M_6C$ carbide phases. These phases are often found in RTAs manufactured by the SPS method at temperatures above 1150-1200°C [66]. The XRD curves from the central layers of all samples revealed the W, Ni, and WO$_2$ phases present in the initial powders as well. It should be noted that the intensities of the WO$_2$ XRD peaks in the sintered samples exceeded the ones in the W@Ni powders. This result indicates the oxidation of the W@Ni powders during the SPS process.

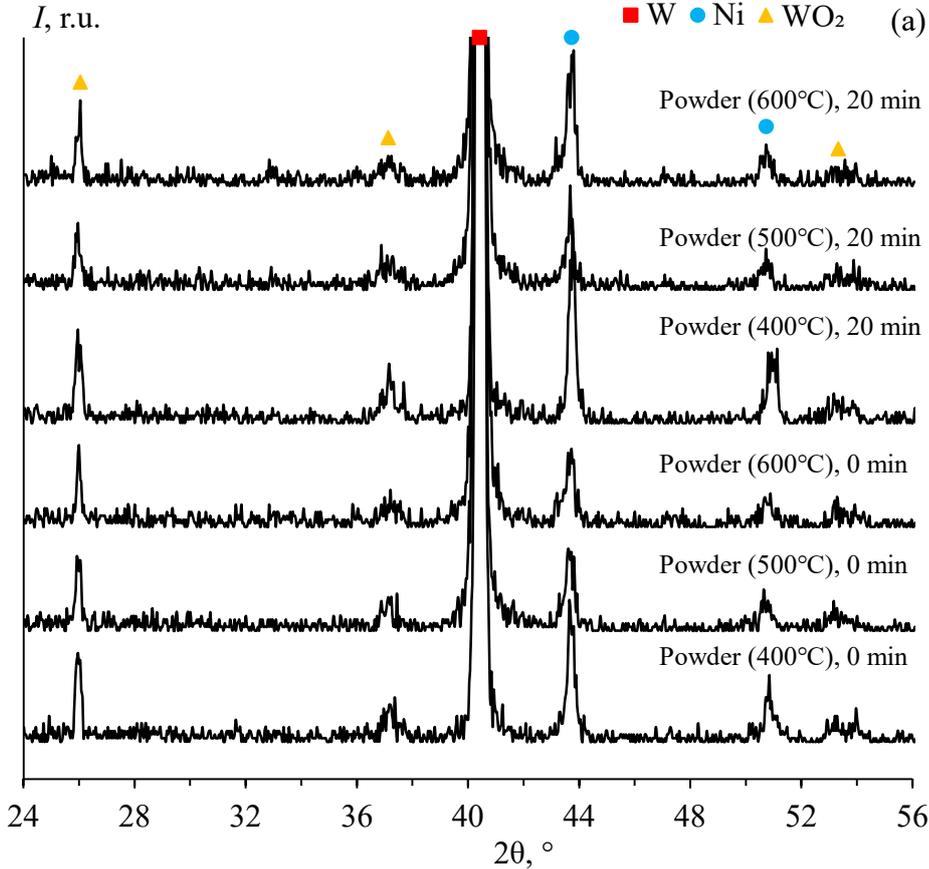

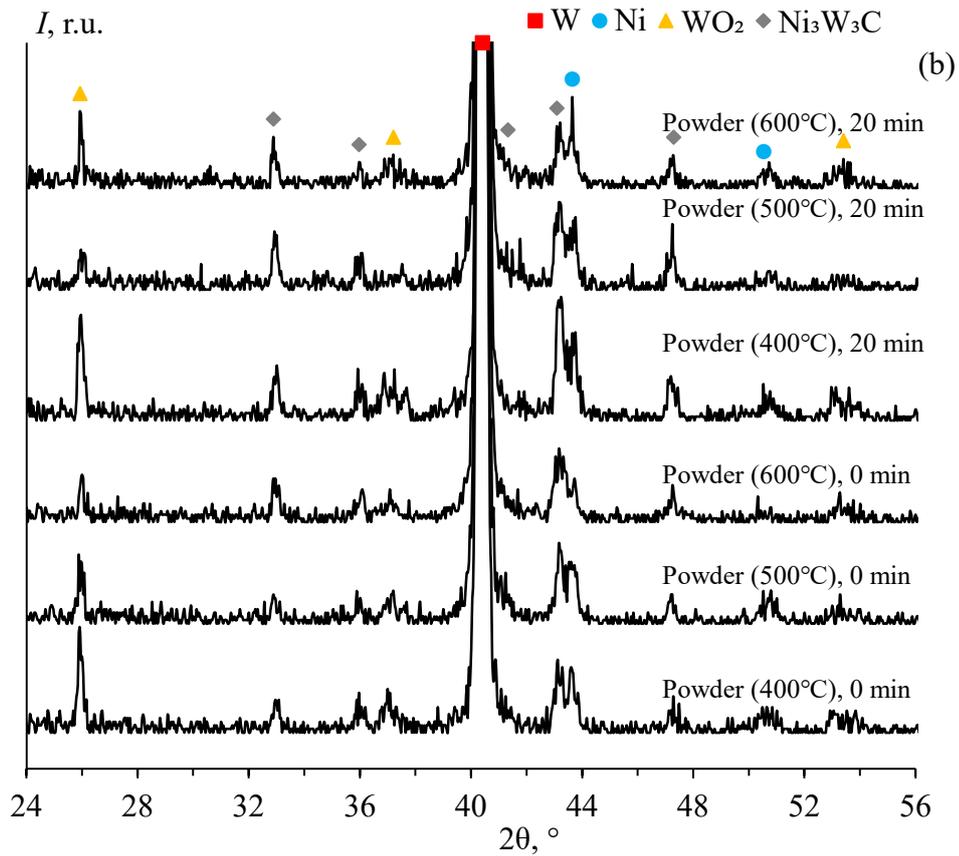

**Fig. 11.** XRD curves from the central (a) and surface (b) zones of the alloy samples #2 sintered from the powders annealed at 400, 500, and 600°C. The isothermal holding time (0 and 20 min) is indicated on the XRD curves

The results of SEM investigations indicated the alloys to have a fairly homogeneous microstructure (Fig. 12). The thickness of the W/Ni interfacial boundaries is very small. Single pores of submicron size are observed on the surface of the grinds, as highlighted with a dotted line in Fig. 12. There were no signatures of abnormal grain growth or formation of large γ-phase particles (Fig. 12). The alloy sintered without isothermal holding from the powders annealed at 400°C had the average grain sizes of ~1-1.5 μm that was close to the initial sizes of the W particles. The Ni based γ-phase was distributed uniformly along the grain boundaries in the W alloy. Fig. 12 shows that increasing the preliminary annealing temperature of W@Ni powders in $H_2$ led to a slight increase in the average W grain sizes to 1.5-2 μm. It should be emphasized that an increase in the average grain size (from 1-1.5 μm after annealing at 400 °C to 1.5-2 μm after annealing at 600°C) only slightly exceeds the measurement error (± 0.5 μm). Additionally, an increase in the isothermal holding time in hydrogen from 0 to 20 min has a smaller effect on grain size than an increase in the annealing temperature in hydrogen from 400 to 600 °C.

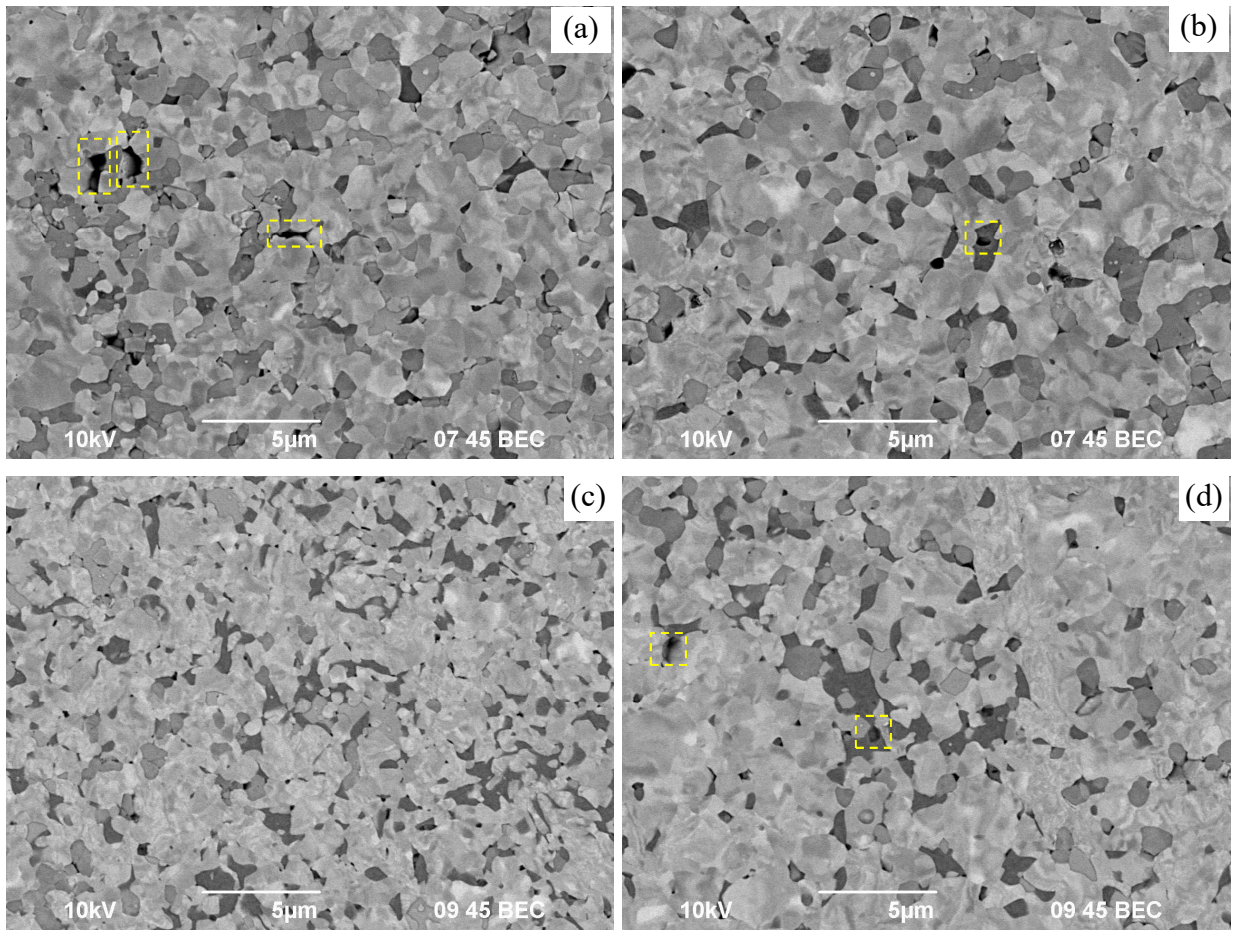

**Fig. 12.** Microstructure of the central layers of alloy samples sintered from the powders annealed at 400°C (a, b) and 600°C (c, d). Sintering without isothermal holding (a, c) and with 20-min holding (b, d). SEM

The average contiguity coefficient $C_{SS}$ in RTAs reaches 0.81-0.90, 0.82-0.88 and 0.76-0.86 in samples produced from powders annealed at 400, 500 and 600°C, respectively. These contiguity coefficient values are relatively low compared to those of coarse-grained tungsten alloys produced by liquid-phase sintering (see [6, 8, 9]). The $C_{SS}$ coefficient determined from the sample fractures was, on average, 0.1 to 0.15 lower than the $C_{SS}$ coefficient calculated from the analysis of metallographic section surfaces.

Fig. 13 presents the results of hardness measurements of the central layers of the alloy samples #2. Table 2 summarizes the results of the studies on the microstructure and mechanical properties of the RTAs of series #2. Fig. 13a shows a decrease in the hardness of the alloys with increasing isothermal holding time due to grain growth. Increasing the annealing temperature of the powders in $H_2$ led to a slight decrease in the hardness. In our opinion, this is due to a decrease in the volume fraction of oxides in the W alloys. The maximum hardness values were observed in the alloys made by SPS from submicron W@Ni powders annealed at 400°C: in the case of sintering without isothermal holding (the hardness is $HV = 5.9 \pm 0.1$ GPa) and with a 20-min isothermal holding ($HV = 5.1 \pm 0.1$ GPa). The hardness of Alloys #2 was comparable to the one of Alloys #1 (Table 1). Severe plastic deformation of the surface occurred during the hardness measurements (when loaded with 10 kg) without crack formation (Fig. 13b). Increasing the load up to 50 kg did not result in the crack formation during the hardness measurements on the surfaces of the W alloy samples. This indicates indirectly an increased fracture toughness of the W alloy series #2.

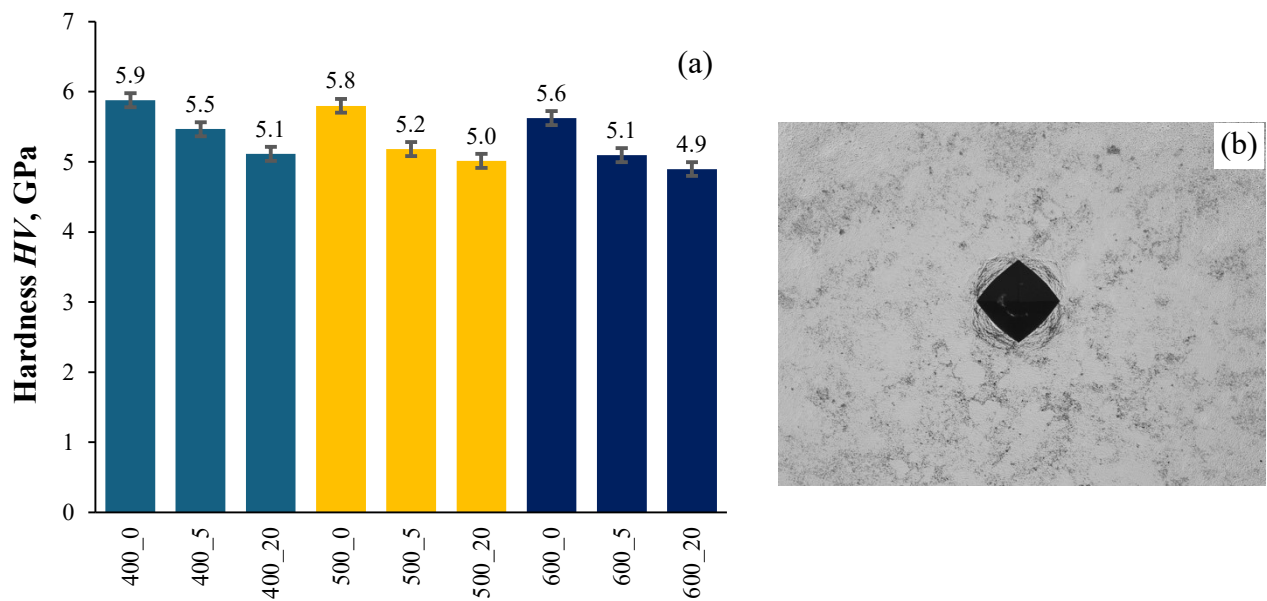

**Fig. 13.** Microhardness of the Alloys #2 (a) and typical view of the imprint from the Vickers pyramid on the sample #2 surface (b). Figure 13a utilizes sample notation in the format '*T_t*' where '*T*' is the hydrogen annealing temperature of the W@Ni powders (in °C) and '*t*' is the time of the isothermal holding at the sintering temperature (1050°C) (in min)

Table 2. Characteristics of RTAs #2 made from W@Ni powders

| $T_{H2}$, °C | SPS regimes | | | | Alloy characteristics | | | |
|---|---|---|---|---|---|---|---|---|
| | $V_h$, °C/min | T, °C | t, min | P, MPa | ρ, % | XRD [1] | d, μm | HV, GPa |
| 400 | 50 | 1050 | 0 | 80 | 99.5 | $WO_2$ | 1 | 5.9 |
| | | | 5 | | 99.9 | | 1.2 | 5.4 |
| | | | 20 | | 99.9 | | 1.2-1.3 | 5.1 |
| 500 | 50 | 1050 | 0 | 80 | 99.7 | $WO_2$ | 1.4-1.5 | 5.8 |
| | | | 5 | | 99.8 | | 1.6 | 5.2 |
| | | | 20 | | 99.9 | | 1.9-2 | 5.0 |
| 600 | 50 | 1050 | 0 | 80 | 100.0 | $WO_2$ | 1.5 | 5.6 |
| | | | 5 | | 100.0 | | 1.7 | 5.1 |
| | | | 20 | | 100.0 | | 2.0 | 4.9 |

[1] The presence of impurity phases in the central layers of samples (based on the XRD analysis).

Figure 14 shows the dependencies of the hardness (HV) on the grain size of Alloys #1 and #2. The $HV - d^{-1/2}$ dependencies for Alloys #2 can be interpolated with good accuracy using straight lines. The absence of a pronounced dependence of hardness on grain size indicates that the main influence on the strength of Alloys #1 is exerted by the characteristics of the nickel phase (phase composition, proportion of impurity phases). The main factor influencing the hardness of Alloys #2 is grain boundary hardening. It can be seen from Fig. 14 that Alloys #1 have a smaller grain size and, at the same time, higher values of hardness. Note that the experimental values of hardness for Alloys #1 are lower than the predicted values, which could be determined by interpolating the $HV - d^{-1/2}$ dependence for Alloys #2. In our opinion, the relatively low values of the contiguity coefficient $C_{SS}$ and the small thickness of the interphase boundaries W/Ni are the primary reasons for the good feasibility of the Hall-Petch ratio in fine-grained Alloys #2 (Fig. 14).

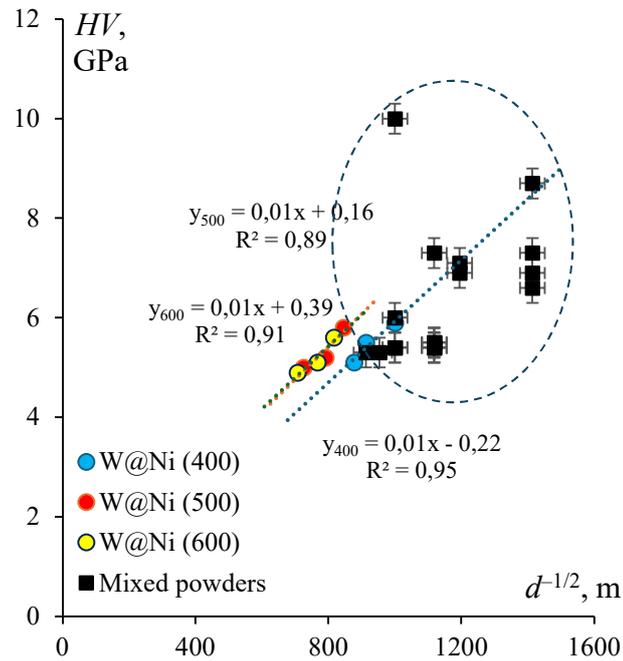

**Fig. 14.** Dependencies of the hardness of Alloys #2 on the grain size. The characteristics of Alloys #1 are indicated by a black square marker, while the characteristics of Alloys #2 are indicated by round markers

### 4. Discussion

First, it is necessary to analyze the SPS mechanisms of the fine W + Ni (Series #1) and W@Ni (Series #2) powders under rapid heating and application of the uniaxial pressure.

As shown above, the temperature curve $L(T)$ has a conventional three-stage character: low-temperature Stage I, when the densification intensity is low, Stage II of intensive shrinkage, and high-temperature Stage III, when the powder shrinkage rate is low again. According to [64], the main mechanism of powder densification at Stage II is the particle rearrangement and their sliding relative to each other under the pressure applied. At Stage III, pore annihilation by diffusion begins that takes place in the grain growth conditions often [64].

In the case of sintering the W-Ni powders, it is important to consider that the initial fine W powders contain an increased concentration of oxygen and $WO_3$ oxides. During annealing in $H_2$, a reaction of Ni with W oxide occurs in the W + Ni + WO$_3$ compositions [62] resulting in the formation of the intermetallic Ni$_4$W. During quasi-stationary heating up to ~1000 °C, a phase transition occurs in the Ni$_4$W intermetallic [67], and the internal stresses are formed at the W/Ni$_4$W interphase boundaries [68]. In our opinion, this is one of the primary origins of the decrease in the density of the W + 10% Ni alloys with increasing SPS temperature from 1000 up to 1075 °C (Table 1). An additional reason for the decrease in the density of Alloys #1 in the presence of Ni$_4$W particles may be the Kirkendall effect, which leads to the formation of pores [69]. In this study, it was demonstrated that the intense diffusion of tungsten in nickel results in diffusion-induced recrystallization at the Ni/Ni$_4$W/W phase boundary, which can take place at ~850-1200 °C.

For the analysis of the mechanical properties, it is also important to note that the phase Ni$_4$W formed at the Ni/W interphase boundaries grows with increasing heating temperature due to the absorption of the ductile Ni phase [70]. This leads to a decrease in the content of the viscous metallic phase Ni in the composition of the Alloys #1, as well as to embrittlement of the W/Ni interfacial boundaries. At the temperatures above 900 °C, the growth of the Ni$_4$W particles at the Ni/W interphase boundaries occurs due to the volume diffusion [70]. Also, note that in the case of high-temperature heating leading to an increased solubility of W in Ni, an additional precipitation of the secondary Ni$_4$W particles may occur during cooling down. These secondary particles would not affect the sintering kinetics but would affect the hardness of the RTAs significantly. The precipitation rate of the Ni$_4$W secondary particles would depend on the sintering temperature, holding time, and heating rate, which affect the solubility of W in Ni.

Secondly, it is necessary to determine the diffusion mechanisms controlling the densification process of submicron W + Ni and W@Ni powders at Stages II and III, which determine the final density of RTAs.

The Young-Cutler model [71] is used to analyze the densification kinetics of the powders during the intensive densification stage (Stage II). This model describes the initial stage of non-isothermal sintering of spherical particles under simultaneous processes of volume and grain boundary diffusion, as well as plastic deformation:

$$\varepsilon^2(\partial\varepsilon/\partial t) = (2{,}63\gamma\Omega D_v\varepsilon/kTd^3) + (0{,}7\gamma\Omega bD_b/kTd^4) + (Ap\varepsilon^2 D/kT), \quad (1)$$

where $\varepsilon$ is the shrinkage, $t$ is the time, $\gamma$ is the free energy, $D_v$ is the volume diffusion coefficient, $D_b$ is the grain boundary diffusion coefficient, $d$ is the grain size, $p$ is the applied pressure (stress), and $D$ is the diffusion coefficient during plastic deformation. The slope of the $\ln(T\partial\varepsilon/\partial T) - T_m/T$ dependence corresponds to the effective activation energy of the sintering process ($mQ_{s2}$) [in $kT_m$], where $m = 1/3$ for grain boundary diffusion, $m = 1/2$ for the volume diffusion, $m = 1$ for the viscous flow (creep) of the material, and $T_m$ is the melting temperature of the material [71, 72]. The efficiency of applying the Young-Cutler model to analyze the sintering mechanisms of the RTAs was demonstrated in [19]. In [19], it was shown that $m = 1$ for the RTAs obtained by mixing the fine powders (W, Ni, and Fe), while $m = 1/3$ during sintering of the mechanically activated nanopowders.

The $\ln(T\partial\varepsilon/\partial T) - T_m/T$ dependencies presented in Fig. 15 exhibit typical two-stage character [71]. The uncertainty of determining the effective activation energy ($Q_{s2}$) was ± 0.3 $kT_m$. When plotting the dependencies $\ln(T\partial\varepsilon/\partial T) - T_m/T$, the melting point of the W + 10% Ni alloy was taken equal to the one of Ni ($T_m = 1726$ K).

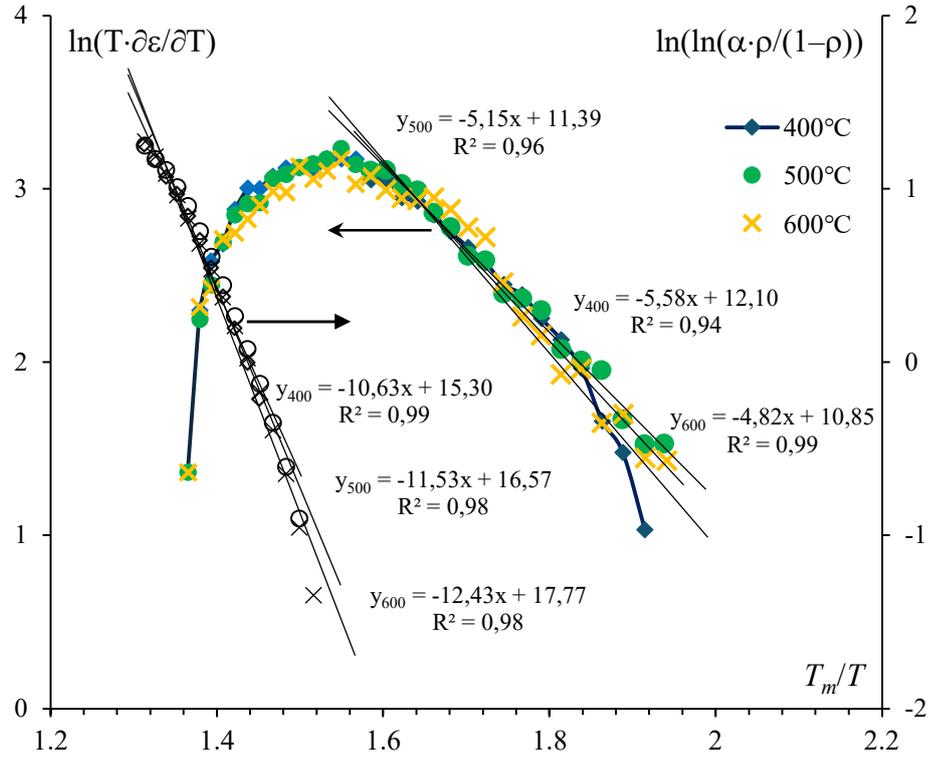

**Fig. 15.** Temperature curves of the Powder #2 shrinkage in the ln($T\partial\varepsilon/\partial T$) – $T_m/T$ and ln(ln($\alpha\cdot\rho/\rho_{th}/(1-\rho/\rho_{th})$)) – $T_m/T$ axes. Calculation of SPS activation energy at Stage II and Stage III

At high heating temperatures, after a short transition stage where the rate of change in ln($T\partial\varepsilon/\partial T$) is very small, the slope of the ln($T\partial\varepsilon/\partial T$) – $T_m/T$ dependence becomes negative (Stage III). This indicates that alternative approaches are necessary to estimate the SPS activation energy at Stage III.

In accordance with [63], the estimation of the activation energy at Stage III can be conducted using the model of diffusion-induced dissolution of pores near the grain boundaries in fine-grained materials. Within the framework of the model [63], the SPS activation energy at Stage III ($Q_{s3}$) can be determined from the slope of the dependence $\rho(T)/\rho_{th}$ in double logarithmic axes ln(ln($\alpha\cdot\rho/\rho_{th}/(1-\rho/\rho_{th})$)) – $T_m/T$ where $\alpha = 0.33$ is the compaction coefficient of the W powder pressing (Fig. 16). The average uncertainty of determining the $Q_{s3}$ was ± 0.5 $kT_m$. The efficiency of applying this model to analyze the SPS activation energy in the THAs was demonstrated in [17, 19].

Using the models [71] and [63], we will analyze the sintering mechanisms of the RTAs at Stage II and Stage III, respectively. The values of SPS effective activation energy of the alloys are presented in Table 3.

*Effect of annealing in hydrogen.* Table 3 shows that for $m = 1$ [19] the value of $mQ_{s2}$ for alloys #1 is close to the activation energy of the grain boundary diffusion in Ni ($Q_{b(Ni)}$ = 115 kJ/mol [73] ~ 8.0 $kT_m$). The result obtained suggests Coble creep to be the main densification mechanism for the Powders #1 at Stage II. This conclusion is consistent with the results of creep studies in RTAs [74]. In our opinion, increased values of the SPS activation energy at Stage II are associated with the presence of the $WO_2$ particles in the initial W + 10% Ni powders #1. It should be noted that $mQ_{s2}$ depends on the temperature of preliminary annealing of the powders in $H_2$ non-monotonically. Annealing at $T_{H2}$ = 700 °C led to a decrease in the $WO_2$ content in the W + 10% Ni powders (Table 1) and to a decrease in the effective SPS activation energy $mQ_{s2}$ from 11.8 down to 8.2 $kT_m$. The SPS activation energy at Stage III decreased from 14.2 down to 11.6 $kT_m$. Further increasing the annealing temperature led to the appearance of a large number of $Ni_4W$ particles (Table 1), which likely hinder the slip of the W particles relative to each other at Stage II and slow down the diffusion annihilation of the pores along the grain boundaries at Stage III. It should be noted that the precipitation of the $Ni_4W$ particles leads to the strengthening of Ni and Ni alloys [75, 76]. The yield stress of Ni alloys with $Ni_4W$ particles at the temperatures above 1000 °C is very high [77] and exceeds the stress, which the sintering of the W + 10% Ni alloys occurs at (40-80 MPa) significantly. Therefore, one can assume that the formation of the $Ni_4W$ particles at elevated annealing temperatures in $H_2$ of powders #1 leads also to an increase in the SPS activation energy (Table 3).

**Table 3.** Activation energy values for SPS of the RTAs

| Synthesis method | $T_{H2}$, °C | SPS regimes | Stage II | | | Stage III |
|---|---|---|---|---|---|---|
| | | | $mQ_{s2}$, $kT_m$ | $m$ | $Q_{s2}$, $kT_m$ | $Q_{s3}$, $kT_m$ |
| #1 | – | | 11.8 | 1 | 11.8 | 14.2 |

| | 700 | 50 °C/min, 80 MPa, 1075 °C | | 8.2 | | 8.2 | 11.6 |
| --- | --- | --- | --- | --- | --- | --- | --- |
| | 900 | | | 7.9 | | 7.9 | 9.2 |
| | 1100 | | | 8.3 | | 8.3 | 17.9 |
| | 1300 | | | 9.7 | | 9.7 | 18.5 |
| | 1050 | 50 °C/min, 1030 °C | 40 MPa | 15.2 | 1 | 15.2 | 30.7 |
| | | | 55 MPa | 14.1 | | 14.1 | 20.6 |
| | | | 60 MPa | 12.8 | | 12.8 | 17.9 |
| | | | 80 MPa | 10.9 | | 10.9 | 13.0 |
| | – | 50 MPa, 1030 °C | 50 °C/min | 13.2 | 1 | 13.2 | 12.5 |
| | | | 100 °C/min | 15.9 | | 15.9 | 14.5 |
| | | | 200 °C/min | 15.8 | | 15.8 | 17.0 |
| | | | 333 °C/min | 18.6 | | 18.6 | 19.5 |
| | | | 500 °C/min | 19.7 | | 19.7 | 20.3 |
| #2 | 400 | 50 °C/min, 70 MPa, 1050 °C | | 5.8 | 1 | 5.8 | 10.6 |
| | 500 | | | 6.1 | | 6.1 | 11.5 |
| | 600 | | | 6.4 | | 6.4 | 12.4 |

*Effect of pressure.* Table 3 illustrates that increasing the pressure applied led to a reduction in the SPS activation energy at Stages II and III. This is associated with easier slip of the W particles relative to each other with increasing pressure applied as well as with an increase in the Coble creep rate of metallic Ni. Accelerated sintering at increased pressure is a well-known effect [64]. It is interesting to note that at low pressure (40 MPa), there is a sharp increase in the SPS activation energy at Stage III. In our opinion, the increase in the SPS activation energy $Q_{s3}$ is due to the high strength of the Ni$_4$W particles and, consequently, the high yield stress of the Ni γ-phase containing the Ni$_4$W particles [75-77]. As a result, the contribution of plastic deformation to the densification of the W alloy at Stage III becomes very small (at σ = 40 MPa), and the diffusion contributes mainly to the densification.

*Effect of heating rate.* The increase in the activation energy with increasing heating rate is the most unexpected and interesting effect. Table 3 shows that increasing $V_h$ from 50 to 500 °C/min led to an increase in the SPS activation energy $Q_{s2}$ from 13.2 to 19.7 $kT_m$ and an increase in $Q_{s3}$ from 12.5 to 20.3 $kT_m$. It is interesting to note also that increasing the heating rate and, hence, the SPS activation energy do not affect the density of the RTAs noticeably (Table 1). This effect is likely due to the non-uniform heating of the W samples at high heating rates [78, 79]. As it is known, at high heating rates, the temperatures of the central parts of the samples are lower than the ones of the samples' surfaces being in contact with the surface of the graphite mold. The lower temperatures of the central parts of the W samples leads to a decrease in the strain rate. Reducing the densification rate of the powders in the formal analysis of the curves $L(T)$ is an equivalent to increasing the SPS activation energy.

*Special features of W@Ni powder sintering.* Table 2 shows that the SPS activation energy of the W@Ni powder is very low ($mQ_{s2} \sim 6\ kT_m$). For $m = 1$, the value of $Q_{s2}$ is much lower than the activation energy of grain boundary diffusion in Ni (~8 $kT_m$ = 115 kJ/mol [73]) and lower than the SPS activation energy for the Powders #1. In our opinion, the primary origin of the low values of $Q_{s2}$ for the W@Ni powders is the absence of the $Ni_4W$ intermetallic particles, which hinder the sintering process of the RTAs. Increasing the temperature of preliminary annealing of the W@Ni powders in $H_2$ from 400 up to 600 °C leads to a slight increase in the SPS activation energy at Stage II ($Q_{s2}$). The scale of increase in $Q_{s2}$ from 5.8 up to 6.4 $kT_m$ with increasing annealing temperature in $H_2$ exceeds the uncertainty of determining the activation energy (± 0.3 $kT_m$) only slightly. Therefore, the effect of this factor can be neglected.

The dependencies of $\ln(\ln(\alpha \cdot \rho/\rho_{th}/(1-\rho/\rho_{th}))) - T_m/T$ for the W@Ni powders at Stage III have typical linear character (Fig. 15). The preliminary annealing temperature does not affect the SPS activation energy $Q_{s3}$ (10.6-12.2 $kT_m$). The SPS activation energy for the W@Ni powders at Stage III exceeds the sintering activation energy at Stage II. The increased values of the SPS activation energy $Q_{s3}$ indicate that grain growth begins at this stage of sintering that corresponds to the results of microstructure investigations (Fig. 12). In the case of grain boundary diffusion, increasing the grain sizes will lead to an increase in the characteristic scale of diffusion mass transfer and, consequently, to an increase in the time required to achieve the theoretical density of the W alloy. It should be noted also that the SPS activation energies $Q_{s3}$ for the W@Ni powders are lower than the ones for powders #1 (Table 2). This result was attributed to lower contents of the $WO_2$ and $Ni_4W$ particles in the W@Ni powders #2 than the ones in the Powders #1.

The following is a brief summary of the results and key differences in the characteristics of fine-grained Alloys #1 and #2 produced by the solid-phase SPS method:

i) The activation energy of sintering of W@Ni composite powders is much lower than the activation energy of solid-phase sintering of mixed W+Ni powders.

ii) Alloys made from W@Ni composite powders have a higher fracture toughness in hardness tests compared to alloys made from mixed W+Ni powders. Alloys made by solid-phase sintering of mixed W+Ni powders are brittle fractured at hardness measurement.

iii) The Hall-Petch equation is satisfied for alloys made from W@Ni powders. The main contribution to the hardness of these alloys is grain boundary hardening. This is due to the small thickness of the W/Ni interphase boundaries in alloys produced from W@Ni powders. There is no clear correlation between hardness and grain size in alloys obtained from mixed W+Ni powders.

**Conclusions**

1. The combined application of chemical-metallurgical deposition of thin Ni layers on the surfaces of the fine W particles and fast SPS allows producing the fine-grained W + 10% wt. Ni alloys manifesting increased relative density and hardness. Optimizing the regimes of preliminary annealing of the W@Ni powders in $H_2$ results in a reduction in the quantity of the $WO_2$ particles in the W alloys and, consequently, in a decrease in the hardness and an increase in the grain sizes. The developed tungsten alloys have ultra-thin W/Ni interfacial boundaries, making then promising for the manufacture of thermo- and plasma-resistant structural elements for ITER.

2. Using Yang-Cutler model and the model of diffusion sintering near the grain boundaries, the SPS activation energies of the W@Ni powders were determined. The activation energy of sintering the W@Ni powders during the intensive shrinkage stage is 5.8-6.4 $kT_m$ and is much lower than the one for the W + Ni powders obtained by mixing. The origin of the abnormally low values of the sintering activation energy for the W@Ni powders is a small fraction of the $WO_2$ particles and $Ni_4W$ intermetallic.

3. The best mechanical properties (the hardness $HV$ = 10.0 GPa) were observed for the W + 10% Ni alloy, manufactured from the unannealed mixed powders at the SPS temperature 1030°C with the heating rate $V_h$ = 50 °C/min at the pressure $P$ = 80 MPa. The samples of alloys made from the unannealed mixed powders exhibited an increased brittleness – the microcracks formed during the hardness measurements with the load of 10 kg. The main contribution to the increase in hardness of the W + 10% Ni alloys is made by the $WO_2$ and $Ni_4W$ particles.

The samples of alloys made from the W@Ni powders annealed in $H_2$ demonstrated the hardness of 4.9-5.9 GPa and increased fracture toughness during indentation. No crack formation was observed during the hardness measurements (at the loads of 10-50 kg). The dependence of hardness on the grain sizes in the alloys sintered from the W@Ni powders obeys the Hall-Petch law. The enhanced mechanical properties of alloys produced from the W@Ni powders are attributed to the thin thickness and extended length of the W/Ni interfacial boundaries.